\documentclass[prl,aps]{revtex4}
\usepackage{psfrag}
\usepackage{graphicx,amsmath,slashed}
\usepackage{dcolumn}
\usepackage{color}
\usepackage{latexsym,amsfonts}
\usepackage{bm}
\usepackage{amssymb}

\begin{document}

\title{Solution of Dirac Equation for Charged and Neutral Fermions
with Anomalous Magnetic Moments in Uniform Magnetic Field}

\author{M. Pitschmann$^{a,b}$, A. N. Ivanov$^c$}
\affiliation{$^a$University of Wisconsin--Madison, Department of
  Physics, 1150 University Avenue, Madison, WI 53706,
  USA}\affiliation{$^b$Physics Division, Argonne National Laboratory,
  Argonne, Illinois 60439, USA}\affiliation{$^c$Atominstitut,
  Technische Universit\"at Wien, Wiedner Hauptstrasse 8-10, A-1040
  Wien, Austria} \email{ivanov@kph.tuwien.ac.at}

\date{\today}

\begin{abstract}
The Dirac equation for charged and neutral fermions with anomalous
magnetic moments is solved in a uniform magnetic field. We find the
relativistic wave functions and energy spectra. In the
non--relativistic limit the wave functions and energy spectra of
charged fermions agree with the known solutions of the Schr\"odinger
equation. 
\end{abstract}

\maketitle

\subsection{1. Introduction}

The quantum field theoretic problem of the motion of charged fermions
with spin $\frac{1}{2}$ in a uniform magnetic field has been solved in
the pioneering papers by Johnson and Lippmann \cite{DMF1}. The wave
functions and energy spectra of the charged fermions have been
obtained as solutions of the Dirac equation at the neglect of the
Pauli term, describing the interaction of the anomalous magnetic
moment of the fermion with a uniform magnetic field \cite{IZ80}.  The
results, obtained in \cite{DMF1}, have been applied to the analysis of
the neutron decay rates in very strong magnetic fields with field
strengths of order of $B \sim 10^{12}\,{\rm T}$
\cite{DMF2}--\cite{DMF5}.  Such strong magnetic fields may exist in
neutron stars and white dwarfs \cite{DMF2}. The solution of the Dirac
equation with the Pauli term for charged and neutral fermions with
spin $\frac{1}{2}$ has been proposed in \cite{DMF6}--\cite{DMF8}. In
\cite{DMF6} only the relativistic energy spectra of
charged fermions with the Pauli energy splitting were found, which are caused by the
interaction of the anomalous magnetic moments of fermions with a uniform
magnetic field. In \cite{DMF7} and \cite{DMF8} the relativistic energy
spectra and the relativistic wave functions of charged fermions (see
\cite{DMF7}) and charged and neutral fermions (see \cite{DMF8}) with
anomalous magnetic moments were found in the ``number of state''
representation, reducing the problem of the motion of fermions with
anomalous magnetic moments in a uniform magnetic field to the harmonic
oscillator problem \cite{DMF9,Landau3}.

In this paper we propose a detailed solution of the Dirac equation
with the Pauli term for fermions with spin $\frac{1}{2}$ and anomalous
magnetic moment $\kappa$. The fermions are either charged $Ze$ with $Z
> 0$ and $Z < 0$ for positively and negatively charged fermions, where
$e$ is the proton charge, or neutral with $Z = 0$, respectively. The
solution of the Dirac equation runs as follows.  We reduce the system
of first order differential equations for coupled {\it up} and {\it
  down} components of the Dirac wave functions to the fourth order
differential equations for the decoupled components, which in turn can
be transformed into a system of second order differential equations
for the decoupled spin {\it up} and spin {\it down} components of the
Dirac wave functions. The solutions of these second order differential
equations give correct relativistic energy spectra, splitted due to
the interaction of the anomalous magnetic moments of fermions with a
uniform magnetic field. The system of first order differential
equations is used for the calculation of the normalisation constants
of the relativistic wave functions. The obtained relativistic
wave functions and energy spectra possess well--known
non--relativistic limits \cite{Landau3}.  We would like to note that
our technique is similar to that, which has been used in
\cite{DMF6}. Nevertheless, our intermediate calculations are much
simpler as well as the derivation of the energy spectra. In addition
we calculate the relativistic wave functions of the charged fermions
that has not been done in \cite{DMF6}. As regards the results obtained
in \cite{DMF7,DMF8}, we would like to accentuate that we solve the
Dirac equation in cylindrical coordinates and define the energy
spectra in dependence on two quantum numbers, namely, the radial
quantum number $n_{\rho} = 0,1,2,\ldots$ and the magnetic quantum
number $m = 0, \pm 1,\pm 2, \ldots$.  According to Mathews
\cite{DMF9}, the ``one--dimensional harmonic oscillator'' description
of relativistic fermions, moving in a uniform magnetic field, cannot
adequately cover the two-dimensional motion of fermions in the plane
orthogonal to the magnetic field.  Thus, following Mathews
\cite{DMF9}, only a two--dimensional description in cylindrical
coordinates with two quantum numbers $(n_{\rho},m)$ can be used for
the correct analysis of relativistic quantum states of fermions in
a uniform magnetic field. Of course, one may also use the principal
quantum number $n$, which is a function of $n_{\rho}$ and $m$, and
the magnetic quantum number $m$. In comparison with the results
obtained in \cite{DMF7,DMF8} we discuss in more detail the dependence
of the relativistic wave functions on the magnetic quantum number,
which testifies an infinite degeneration of the energy levels of
fermions in the ``one--dimensional harmonic oscillator''
representation \cite{DMF9} in a uniform magnetic field. It also plays
an important role for applications of the obtained results, for
example to the neutron decay that requires the calculation
matrix elements by integrating over the spatial coordinates and
summation over quantum numbers.

The paper is organised as follows. In section 2 we give a detailed
solution of the Dirac equation for charged fermions with spin
$\frac{1}{2}$ and an anomalous magnetic moment $\kappa$. In section 3
we propose the solution of the Dirac equation for neutral fermions
with spin $\frac{1}{2}$ and an anomalous magnetic moment $\kappa$. In
section 4 we analyse the non--relativistic limit of the relativistic
wave functions and energy spectra, obtained in sections 2 and 3. In
the Conclusion we discuss the obtained results.

\subsection{2. Solution of Dirac equation for charged fermions with 
anomalous magnetic moments in uniform magnetic field}

Let a fermion with mass $M$, spin $\frac{1}{2}$, electric charge $Ze$
and an anomalous magnetic moment $\kappa$ move in a uniform magnetic
field $\vec{B} = B\,\vec{e}_z$, directed along the $z$--axis, which
coincides with the quantisation axis of the fermion spin. The Dirac
equation takes the form
\cite{IZ80}
\begin{eqnarray}\label{label1}
  \Big(\gamma^\mu(i\partial_\mu - ZeA_\mu(x)) +
  \lambda \,\sigma_{\mu\nu}F^{\mu\nu}(x) - M\Big)\psi(x) = 0,
\end{eqnarray}
where $x = (t, \vec{r}\,)$, $\gamma^{\mu} = (\gamma^0,\vec{\gamma}\,)
= (\beta, \beta\,\vec{\alpha}\,)$ and $\sigma_{\mu\nu} =
\frac{i}{2}(\gamma_{\mu}\gamma_{\nu} - \gamma_{\nu}\gamma_{\mu})$ are
Dirac matrices in the Dirac representation \cite{IZ80}. Then, $\lambda
\,\sigma_{\mu\nu}F^{\mu\nu}(x)$ is the Pauli term, where $\lambda =
\kappa Ze/4M$, $F^{\mu\nu}(x) = \partial^{\mu}A^{\nu}(x) -
\partial^{\nu}A^{\mu}(x)$ is the electromagnetic field strength and
$A^{\mu}(x) = (A^0(x),\vec{A}(x))$ is the electromagnetic
potential. For a uniform magnetic field, where $A^0(x) = 0$ and
$\vec{A}(0,\vec{r}\,) = \frac{1}{2}(\vec{B}\times \vec{r}\,)$,
Eq.(\ref{label1}) can be transcribed into the from
\begin{eqnarray}\label{label2}
  i\frac{\partial}{\partial
  t}\,\psi(x)=\Big(\vec{\alpha}\cdot\vec{\pi} -
  2\beta\lambda \,\vec\Sigma\cdot\vec B + \beta M\Big)\psi(x).
\end{eqnarray}
Here $\vec{\pi} = - i\vec\nabla - Ze\vec{A}$ is the operator of the
canonical momentum of the fermion with charge $Ze$ in a magnetic field
$\vec{B}$ and $ 2\beta\lambda \,\vec\Sigma\cdot\vec B$ is the Pauli
term, where $\vec{\Sigma} = \gamma^0\vec{\gamma}\gamma^5$ is a
diagonal matrix ${\rm diag}(\vec{\sigma},\vec{\sigma}\,)$, the
elements of which $\vec{\sigma} = (\sigma_x,\sigma_y,\sigma_z)$ are
Pauli $2\times 2$ matrices \cite{IZ80}. Since we search for stationary
solutions, we take the wave function of the fermion in the following
form
\begin{eqnarray}\label{label3}
  \psi(x) = e^{-iEt} \left(\begin{array}{c} \varphi(\vec{r}\,) \\
  \chi(\vec{r}\,) \\
  \end{array}\right),
\end{eqnarray}
where $\vec{r} = (\vec{r}_{\perp}, z) = (x, y, z)$ and
 $\varphi(\vec{r}\,)$ and $\chi(\vec{r}\,)$ are two--component wave
 functions, obeying the equations
\begin{eqnarray}\label{label4}
  &&\Big(E - M + 2\lambda \,\vec\sigma\cdot\vec B\Big)\varphi(\vec{r}\,) =
  \vec\sigma\cdot\vec\pi\,\chi(\vec{r}\,), \nonumber\\ &&\Big(E + M -
  2\lambda \,\vec\sigma\cdot\vec B\Big)\chi(\vec{r}\,) =
  \vec\sigma\cdot\vec\pi\,\varphi(\vec{r}\,).
\end{eqnarray}
Taking into account that the magnetic field is directed along the
$z$--axis, we can rewrite Eq.(\ref{label4}) as follows
\begin{eqnarray}\label{label5}
  \hspace{-0.3in}&&\Big(E - M + 2\lambda \,\sigma_z
  B\Big)\varphi(\vec{r}\,) = \Big(\sigma_+\pi_- + \sigma_-\pi_+ +
  \sigma_z\pi_z\Big)\chi(\vec{r}\,), \nonumber\\
 \hspace{-0.3in}&&\Big(E + M - 2\lambda \,\sigma_z
  B\Big)\chi(\vec{r}\,) = \Big(\sigma_+\pi_- + \sigma_-\pi_+ +
  \sigma_z\pi_z\Big)\varphi(\vec{r}\,),
\end{eqnarray}
where $\pi_\pm = \pi_x \pm i\pi_y$ and $\pi_z = -
i\partial_z$. Specifying the direction of the fermion spin ``up'' and
``down'' as $\uparrow$ and $\downarrow$, the two--component spinor
wave functions take the form
\begin{eqnarray}\label{label6}
  \varphi(\vec{r}\,) = \left(
  \begin{array}{c}
  \varphi_{\uparrow}(\vec{r}\,) \\
  \varphi_{\downarrow}(\vec{r}\,) \\
  \end{array}
  \right)\quad,\quad \chi(\vec{r}\,)=\left(
  \begin{array}{c}
  \chi_{\uparrow}(\vec{r}\,) \\
  \chi_{\downarrow}(\vec{r}\,) \\
  \end{array}
  \right),
\end{eqnarray}
where $\varphi_{\uparrow,\downarrow}$ and $\chi_{\uparrow,\downarrow}$
are eigenfunctions of the $\sigma_z$--operator, i.e. $\sigma_z
\varphi_\uparrow = +\varphi_\uparrow$, $\sigma_z
\varphi_\downarrow = -\varphi_\downarrow$, $\sigma_z \chi_\uparrow =
+\chi_\uparrow$ and $\sigma_z \chi_\downarrow =
-\chi_\downarrow$. Since the
$p_z$--component of the fermion is conserved, the wave functions
$\varphi_{\uparrow,\downarrow}(\vec{r}\,)$ and
$\chi_{\uparrow,\downarrow}(\vec{r}\,)$ take the form
$\varphi_{\uparrow,\downarrow}(\vec{r}\,) =
\varphi_{\uparrow,\downarrow}(\vec{r}_{\perp}\,)\,e^{ip_zz}$ and
$\chi_{\uparrow,\downarrow}(\vec{r}\,) =
\chi_{\uparrow,\downarrow}(\vec{r}_{\perp}\,)\,e^{ip_zz}$,
respectively.  Substituting the wave functions Eq.(\ref{label6}) in
Eq.(\ref{label5}) we find for the wave functions
$\varphi_{\uparrow,\downarrow}(\vec{r}_{\perp}\,)$ and
$\chi_{\uparrow,\downarrow}(\vec{r}_{\perp}\,)$ the following
system of first order differential equations
\begin{eqnarray}\label{label7}
  &&\pi_+\varphi_{\uparrow}(\vec{r}_{\perp}) = \Big(E + M + 2\lambda 
  B\Big)\chi_{\downarrow}(\vec{r}_{\perp}) +
  p_z\,\varphi_{\downarrow}(\vec{r}_{\perp}),\nonumber\\
 &&\pi_-\varphi_{\downarrow}(\vec{r}_{\perp}) = \Big(E + M - 2\lambda 
  B\Big)\chi_{\uparrow}(\vec{r}_{\perp}) -
  p_z\,\varphi_{\uparrow}(\vec{r}_{\perp}), \nonumber\\
 &&\pi_+\chi_{\uparrow}(\vec{r}_{\perp}) = \Big(E - M - 2\lambda 
  B\Big)\varphi_{\downarrow}(\vec{r}_{\perp}) +
  p_z\,\chi_{\downarrow}(\vec{r}_{\perp}), \nonumber\\
 &&\pi_-\chi_{\downarrow}(\vec{r}_{\perp}) = \Big(E - M + 2\lambda 
  B\Big)\varphi_{\uparrow}(\vec{r}_{\perp}) -
  p_z\,\chi_{\uparrow}(\vec{r}_{\perp}).
\end{eqnarray}
Acting on these equations with the operators $\pi_{\pm}$ and using the
commutation relation $[\pi_+,\pi_-] = 2 Z e B$ we arrive at the system
of second order differential equations
\begin{eqnarray}\label{label8}
\hspace{-0.3in}\Big(\frac{1}{2}\,\{\pi_+,\pi_-\} - (E + 2\lambda 
  B)^2 + M^2 - ZeB + p_z^2\Big) \varphi_{\uparrow}(\vec{r}_{\perp})
  &=& - \,4\lambda  B\,p_z \chi_{\uparrow}(\vec{r}_{\perp}),\nonumber\\
\hspace{-0.3in}\Big(\frac{1}{2}\,\{\pi_+,\pi_-\} - (E - 2\lambda 
  B)^2 + M^2 + ZeB + p_z^2\Big)\varphi_{\downarrow}(\vec{r}_{\perp}) &=& -
  \,4\lambda  B\,p_z \chi_{\downarrow}(\vec{r}_{\perp}), \nonumber\\
 \hspace{-0.3in}\Big(\frac{1}{2}\,\{\pi_+,\pi_-\} - (E - 2\lambda 
  B)^2 + M^2 - ZeB + p_z^2\Big)\chi_{\uparrow}(\vec{r}_{\perp}) &=& + \,4
  \lambda  B\,p_z \varphi_{\uparrow}(\vec{r}_{\perp}), \nonumber\\
 \hspace{-0.3in}\Big(\frac{1}{2}\,\{\pi_+,\pi_-\} - (E + 2\lambda 
  B)^2 + M^2 + ZeB + p_z^2\Big)\chi_{\downarrow}(\vec{r}_{\perp}) &=&
  + \,4 \lambda  B\,p_z \varphi_{\downarrow}(\vec{r}_{\perp}),
\end{eqnarray}
where the components $\uparrow$ and $\downarrow$ of the wave functions
are decoupled.  

The disentangled components $\varphi_{\uparrow}(\vec{r}_{\perp}),
\chi_{\uparrow}(\vec{r}_{\perp})$ and
$\varphi_{\downarrow}(\vec{r}_{\perp}),
\chi_{\downarrow}(\vec{r}_{\perp})$ of the fermion wave function are
described by the fourth order differential equations
\begin{eqnarray}\label{label9}
 \hspace{-0.3in}&&\Big(\frac{1}{2}\,\{\pi_+,\pi_-\} - E_\perp^2 + M^2
 - ZeB - 4\lambda^2 B^2\Big)^2 \Phi_{\uparrow}(\vec{r}_{\perp}) =
 16\lambda^2 B^2E_\perp^2\,\Phi_{\uparrow}(\vec{r}_{\perp}),\nonumber\\
\hspace{-0.3in}&&\Big(\frac{1}{2}\,\{\pi_+,\pi_-\} - E_\perp^2 + M^2 +
  ZeB - 4\lambda^2 B^2\Big)^2 \Phi_{\downarrow}(\vec{r}_{\perp}) =
  16\lambda^2 B^2E_\perp^2\,\Phi_{\downarrow}(\vec{r}_{\perp}),
\end{eqnarray}
where $\Phi_{\uparrow,\downarrow}(\vec{r}_{\perp}) =
\varphi_{\uparrow,\downarrow}(\vec{r}_{\perp})$ or
$\chi_{\uparrow,\downarrow}(\vec{r}_{\perp})$ and $E_\perp^2 = E^2 - p_z^2$.

The fourth order differential equations Eq.(\ref{label9}) can be
transmitted into the second order differential equations 
\begin{eqnarray}\label{label10}
\hspace{-0.3in}&&\Big(\frac{1}{2}\,\{\pi_+,\pi_-\} - E_\perp^2 + M^2 -
ZeB - 4\lambda^2 B^2\Big) \Phi^{(\pm)}_{\uparrow}(\vec{r}_{\perp}) = \pm
4\lambda  B E_{\perp} \Phi^{(\pm)}_{\uparrow}(\vec{r}_{\perp}) ,\nonumber\\
 \hspace{-0.3in}&&\Big(\frac{1}{2}\,\{\pi_+,\pi_-\} - E_\perp^2 + M^2
 + ZeB - 4\lambda^2
 B^2\Big)\Phi^{(\pm)}_{\downarrow}(\vec{r}_{\perp}) = \pm 4\lambda 
 B E_\perp \Phi^{(\pm)}_{\downarrow}(\vec{r}_{\perp}),
\end{eqnarray} 
which can be rewritten in the following form
\begin{eqnarray}\label{label11}
\hspace{-0.3in}&&\Big(\frac{1}{2}\,\{\pi_+,\pi_-\} - (E_\perp \pm
2\lambda  B)^2 + M^2 -
ZeB\Big)\Phi^{(\pm)}_{\uparrow}(\vec{r}_{\perp}) = 0,\nonumber\\
 \hspace{-0.3in}&&\Big(\frac{1}{2}\,\{\pi_+,\pi_-\} - (E_\perp \pm
  2\lambda  B)^2 + M^2 +
  ZeB\Big)\Phi^{(\pm)}_{\downarrow}(\vec{r}_{\perp})= 0.
\end{eqnarray} 
We obtain the solutions of these equations in
cylindrical coordinates $\vec{r} = (\rho, \phi, z)$ \cite{DMF9}, where
they read (see also \cite{DMF9})
\begin{eqnarray}\label{label12}
  \hspace{-0.3in}&&\Big[\frac{1}{\rho}\frac{\partial}{\partial
  \rho}\Big(\rho \frac{\partial}{\partial \rho}\Big) +
  \frac{1}{\rho^2}\,\frac{\partial^2}{\partial\phi^2} - ZeB\,
  i\frac{\partial}{\partial\phi} - \frac{Z^2e^2B^2\rho^2}{4} +
  (E_\perp \pm 2\lambda  B)^2 - M^2 + ZeB\Big]
  \Phi^{(\pm)}_{\uparrow}(\vec{r}_{\perp}) = 0,\nonumber\\
\hspace{-0.3in}&&\Big[\frac{1}{\rho}\frac{\partial}{\partial
  \rho}\Big(\rho \frac{\partial}{\partial \rho}\Big) +
  \frac{1}{\rho^2}\,\frac{\partial^2}{\partial\phi^2} - ZeB\,
  i\frac{\partial}{\partial\phi} - \frac{Z^2e^2B^2\rho^2}{4} +
  (E_\perp \pm 2\lambda B)^2 - M^2 -
  ZeB\Big]\Phi^{(\pm)}_{\downarrow}(\vec{r}_{\perp}) = 0.
\end{eqnarray}
We obtain the unnormalised solutions
$f_{n_{\rho},m}(\vec{r}_{\perp})$ of these equations in the form (see also \cite{Lebedew,Landau3})
\begin{eqnarray}\label{label13}
 f_{n_{\rho},m}(\vec{r}_{\perp}) = \rho^{|m|}\,
 L^{|m|}_{n_{\rho}}\Big(\frac{|Z|eB\rho^2}{2}\Big)\,e^{\textstyle -
 \frac{|Z|eB\rho^2}{4}}\,e^{i m \phi},
\end{eqnarray}
where $n_{\rho} = 0,1,2,\ldots$ and $m = 0, \pm 1, \pm 2, \ldots$ are
the radial and magnetic quantum numbers \cite{Landau3}, respectively,
$L^{|m|}_{n_{\rho}}(|Z|eB\rho^2/2)$ are the generalised Laguerre
polynomials, defined by  \cite{Lebedew}
\begin{eqnarray}\label{label14}
 L^{|m|}_{n_{\rho}}(\xi) = e^{\xi}\,\frac{\xi^{- |m|}}{\Gamma(n_{\rho} +
   1)}\frac{d^{n_{\rho}}}{d\xi^{n_{\rho}}}\Big( e^{- \xi}
 \xi^{n_{\rho} + |m|}\Big),
\end{eqnarray}
where $\xi = |Z|eB\rho^2/2$ for a fermion in a uniform magnetic field
and $\Gamma(n_{\rho} + 1)$ is the Euler $\Gamma$--function
$\Gamma(n_{\rho} + 1) = n_{\rho}!$. According to the definition of the
generalised Laguerre polynomials Eq.(\ref{label14}), the polynomials
vanish for $n_{\rho} \le -1$. 

The energy spectra for the $ \Phi^{(\pm)}_{\uparrow}(\vec{r}_{\perp})$
and $\Phi^{(\pm)}_{\downarrow}(\vec{r}_{\perp})$ components of the
fermion wave functions are
\begin{eqnarray}\label{label15}
  \hspace{-0.3in}E^{(\pm, Z > 0)}_{\perp\uparrow,n_{\rho},m} &=& \sqrt{M^2 +
ZeB\,(2n_{\rho} + |m| - m )} \mp  2\lambda  B,\nonumber\\
  \hspace{-0.3in}E^{(\pm,Z > 0)}_{\perp\downarrow,n_{\rho},m} &=&
  \sqrt{M^2 + ZeB\,(2(n_{\rho} + 1) + |m| - m)} \mp 2\lambda  B,
\end{eqnarray}
and
\begin{eqnarray}\label{label16}
  \hspace{-0.3in}E^{(\pm, Z < 0)}_{\perp\uparrow,n_{·\rho},m} &=& \sqrt{M^2 +
    |Z|eB\,(2(n_{\rho} + 1) + |m| + m )} \mp  2\lambda  B,\nonumber\\
  \hspace{-0.3in}E^{(\pm, Z < 0)}_{\perp\downarrow,n_{·\rho},m} &=&
  \sqrt{M^2 + |Z|eB\,(2 n_{\rho} + |m| + m)} \mp 2\lambda B,
\end{eqnarray}
for positively and negatively charged fermions, respectively. 

In order to define the Dirac wave functions we have to take into
account that
$\varphi^{(\pm)}_{\uparrow,\downarrow}(\vec{r}_{\perp})$ and
$\chi^{(\pm)}_{\uparrow,\downarrow}(\vec{r}_{\perp})$ satisfy the
system of the first order differential equations
Eq.(\ref{label7}). For this aim we use the following relations
\begin{eqnarray}\label{label17}
  \hspace{-0.3in}\pi_+ f_{n_{\rho},m}(\vec{r}_{\perp}) &=&
  -\,i\,\rho^{|m| - 1}\,e^{\textstyle
  -\,\frac{|Z|eB}{4}\,\rho^2}\,e^{i(m + 1)\phi}\,\Big[(|m| -
  m)L^{|m|}_{n_{\rho}} - \frac{(|Z| - Z)eB}{2}\,\rho^2 L^{|m|}_{n_{\rho}} +
  \rho\frac{d L^{|m|}_{n_{\rho}}}{d\rho}\Big],\nonumber\\
\hspace{-0.3in}\pi_- f_{n_{\rho},m}(\vec{r}_{\perp}) &=&
  -\,i\,\rho^{|m| - 1}\,e^{\textstyle
  -\,\frac{|Z|eB}{4}\,\rho^2}\,e^{i(m - 1)\phi}\,\Big[(|m| +
  m)L^{|m|}_{n_{\rho}} - \frac{(|Z| + Z)eB}{2}\,\rho^2
  L^{|m|}_{n_{\rho}} + \rho\frac{d L^{|m|}_{n_{\rho}}}{d\rho}\Big].
\end{eqnarray}
For positively charged fermions $Z > 0$ and the states with a magnetic
number $m \ge 0$ we get the relations
\begin{eqnarray}\label{label18}
  \hspace{-0.3in}\pi_+ f_{n_{\rho},m}(\vec{r}_{\perp}) &=&+\,i
  ZeB\,f_{n_{\rho}-1,m+1}(\vec{r}_{\perp}),\nonumber\\
 \hspace{-0.3in}\pi_- f_{n_{\rho} - 1,m + 1}(\vec{r}_{\perp})
 &=&-\,2i\,n_{\rho}\,f_{n_{\rho},m}(\vec{r}_{\perp}).
\end{eqnarray}
In turn for fermions with $Z > 0$ and the states with a magnetic
number $m < 0$ the corresponding relations are
\begin{eqnarray}\label{label19}
  \hspace{-0.3in}\pi_+ f_{n_{\rho},m}(\vec{r}_{\perp}) &=&-\,2i\,(n_{\rho} +
  |m|)\,f_{n_{\rho}, m + 1}(\vec{r}_{\perp}),\nonumber\\
 \hspace{-0.3in}\pi_- f_{n_{\rho},m + 1}(\vec{r}_{\perp})
 &=&+\,i\,ZeB\,f_{n_{\rho},m}(\vec{r}_{\perp}).
\end{eqnarray}
For negatively charged fermions $Z < 0$ and the states with magnetic number $m \ge 0$ we obtain the relations
\begin{eqnarray}\label{label20}
  \hspace{-0.3in}\pi_+ f_{n_{\rho},m}(\vec{r}_{\perp}) &=&+\,i
  |Z|eB\,f_{n_{\rho},m+1}(\vec{r}_{\perp}),\nonumber\\
 \hspace{-0.3in}\pi_- f_{n_{\rho},m + 1}(\vec{r}_{\perp})
 &=&-\,2i\,(n_{\rho} + m + 1)\,f_{n_{\rho},m}(\vec{r}_{\perp}),
\end{eqnarray}
and for fermions with $Z < 0$ and magnetic number $m < 0$ we finally obtain
\begin{eqnarray}\label{label21}
  \hspace{-0.3in}\pi_+ f_{n_{\rho},m}(\vec{r}_{\perp}) &=&-\,2i\,(n_{\rho} +
  1)\,f_{n_{\rho} + 1, m + 1}(\vec{r}_{\perp}),\nonumber\\
 \hspace{-0.3in}\pi_- f_{n_{\rho} + 1, m + 1}(\vec{r}_{\perp})
 &=&+\,i\,|Z|eB\,f_{n_{\rho},m}(\vec{r}_{\perp}).
\end{eqnarray}
As a result the Dirac wave
functions of charged fermions with anomalous magnetic moments are
defined by
\begin{eqnarray}\label{label22}
 \hspace{-0.3in} \psi^{(\pm, Z > 0)}_{n_{\rho},m \ge 0, p_z}(x) =
  \left(\begin{array}{c} A_{1}\,f_{n_{\rho},m}(\vec{r}_{\perp})\\
  A_2\,f_{n_{\rho}-1,m + 1}(\vec{r}_{\perp})\\
  A_{3}\,f_{n_{\rho},m}(\vec{r}_{\perp}) \\ A_4\,f_{n_{\rho}-1,m +
  1}(\vec{r}_{\perp})
  \end{array}\right)\, 
e^{-iE^{(\pm, Z > 0)}_{n_{\rho},m \ge 0} t + ip_z z}\;,\; \psi^{(\pm, Z >
  0)}_{n_{\rho},m < 0, p_z}(x) = \left(\begin{array}{c}
  B_1\,f_{n_{\rho},m}(\vec{r}_{\perp}) \\ B_2\,f_{n_{\rho},m + 1}(\vec{r}_{\perp})
  \\ B_3\,f_{n_{\rho},m}(\vec{r}_{\perp}) \\ B_4 \,f_{n_{\rho},m +
  1}(\vec{r}_{\perp})
  \end{array}\right) \, 
e^{-iE^{(\pm, Z > 0)}_{n_{\rho},m < 0} t + ip_z z},
\end{eqnarray}  
and
\begin{eqnarray}\label{label23} 
 \hspace{-0.3in}\psi^{(\pm, Z < 0)}_{n_{\rho},m \ge 0}(x)=
 \left(\begin{array}{c} C_1\,f_{n_{\rho}-1,m}(\vec{r}_{\perp})\\ 
 C_2\,f_{n_{\rho}-1,m +
     1}(\vec{r}_{\perp}) \\ 
     C_3\,f_{n_{\rho}-1,m}(\vec{r}_{\perp})\\ 
     C_4\,f_{n_{\rho}-1,m + 1}(\vec{r}_{\perp})\end{array}\right)\, e^{-iE^{(\pm, Z <
     0)}_{n_{\rho},m \ge 0} t + ip_z z}\;,\; \psi^{(\pm, Z <
   0)}_{n_{\rho},m < 0}(x) = \left(\begin{array}{c} D_1\,f_{n_{\rho}-1,m}(\vec{r}_{\perp}) \\ 
   D_2\,f_{n_{\rho},m +
     1}(\vec{r}_{\perp}) \\ 
     D_3\,f_{n_{\rho}-1,m}(\vec{r}_{\perp})\\ 
     D_4\,f_{n_{\rho},m + 1}(\vec{r}_{\perp})
  \end{array}\right)\,
e^{-iE^{(\pm, Z < 0)}_{n_{\rho},m < 0} t + ip_z z},
\end{eqnarray}
where $n_{\rho} = 0,1,\ldots$ and $m = 0,\pm 1,\pm 2,\ldots$. The
relativistic energy spectra are given for $Z > 0$ by 
\begin{eqnarray}\label{label24} 
\hspace{-0.3in}&&E^{(\pm, Z > 0)}_{n_{\rho},m} = \sqrt{(\sqrt{M^2 +
|Z|eB\,(2n_{\rho} + |m| - m)} \mp  2\lambda B)^2 + p^2_z},
\end{eqnarray}
and for $Z < 0$ by
\begin{eqnarray}\label{label25} 
\hspace{-0.3in}&&E^{(\pm, Z < 0)}_{n_{\rho},m} =
\sqrt{(\sqrt{M^2 + |Z|eB\,(2n_{\rho} + |m| + m)} \mp 2\lambda B)^2 +
  p^2_z}.
\end{eqnarray}
The Dirac wave functions are normalised by
\begin{eqnarray}\label{label26}  
  \hspace{-0.3in}\int d^3x\,\psi^{(\sigma', Z)\dagger}_{n'_{\rho},m',p'_z}(x)
  \psi^{(\sigma, Z)}_{n_{\rho},m, p_z}(x) = 2 E^{(\sigma,
  Z)}_{n_{\rho},m}\,(2\pi)^3\,\delta_{n_{\rho}n'_{\rho}}\delta_{m m'}\,\delta(p_z -
  p_z')\,\delta_{\sigma'\sigma},
\end{eqnarray}
where $\sigma = \pm$.  For the calculation of the normalisation
constants we use the following relation
\begin{eqnarray}\label{label27}  
 \hspace{-0.3in}\int
 d^2x\,f_{n'_{\rho},m'}^*(\vec{r}_{\perp})f_{n_{\rho},m}(\vec{r}_{\perp})
 = \frac{\pi(n_{\rho} + |m|)!2^{|m| + 1}}{n_{\rho}!(|Z|eB)^{|m| +
     1}}\, \delta_{n_{\rho} n'_{\rho}}\delta_{m m'}.
\end{eqnarray}
Substituting Eq.(\ref{label22}) and Eq.(\ref{label23}) into
Eq.(\ref{label7}) we arrive at a system of algebraical equations for
the normalisation constants $A_i$, $B_i$, $C_i$ and $D_i$ for $i =
1,2,3,4$.  Solving these equations together with the normalisation
condition Eq.(\ref{label26}) and using Eq.(\ref{label27}) we obtain
the following normalised Dirac wave functions of positively charged
($Z > 0$) fermions
\begin{eqnarray}\label{label28}
 \hspace{-0.3in}&& \psi^{(\pm, Z > 0)}_{n_{\rho},m \ge 0, p_z}(x) =
 \sqrt{\frac{\pi
     (n_{\rho}-1)!(ZeB)^{m+2}}{2^{m+1}(n_{\rho}+m)!}}\frac{1}{\sqrt{(E^{(\pm,Z
       > 0)}_{n_{\rho},m \ge 0} \pm E^{(\pm,Z > 0)}_{\perp,n_{\rho},m
       \ge 0})(E^{(\pm,Z > 0)}_{\perp,n_{\rho},m \ge 0} \pm 2\lambda
     B) (E^{(\pm,Z > 0)}_{\perp,n_{\rho},m \ge 0} \pm 2\lambda B \mp
     M)}}\nonumber\\
 \hspace{-0.3in}&& \times\left(\begin{array}{c} 2 i n_{\rho}\,(E^{(\pm,Z >
 0)}_{n_{\rho},m \ge 0} \pm E^{(\pm,Z > 0)}_{\perp,n_{\rho},m \ge
 0})\,f_{n_{\rho},m}(\vec{r}_{\perp})\\ \pm p_z (E^{(\pm,Z > 0)}_{\perp,n_{\rho},m
 \ge 0} \pm 2 \lambda B \mp M)\,f_{n_{\rho}-1,m + 1}(\vec{r}_{\perp})\\ 2 i
 n_{\rho} p_z\,f_{n_{\rho},m}(\vec{r}_{\perp}) \\ \mp (E^{(\pm,Z > 0)}_{\perp,n_{\rho},m
 \ge 0} \pm 2\lambda B \mp M)\,(E^{(\pm,Z > 0)}_{n_{\rho},m \ge 0} \pm
 E^{(\pm,Z > 0)}_{\perp,n_{\rho},m \ge 0})\,f_{n_{\rho}-1,m + 1}(\vec{r}_{\perp})
  \end{array}\right)\, 
e^{-iE^{(\pm, Z > 0)}_{n_{\rho},m \ge 0} t + ip_z z},
\end{eqnarray}
and 
\begin{eqnarray}\label{label29}
\hspace{-0.3in}&& \psi^{(\pm, Z > 0)}_{n_{\rho},m < 0, p_z}(x) =
\sqrt{\frac{\pi n_{\rho}!(ZeB)^{|m|}}{2^{|m| -
      1}(n_{\rho}+|m|-1)!}}\frac{1}{\sqrt{(E^{(\pm,Z > 0)}_{n_{\rho},m
      < 0} \pm E^{(\pm,Z > 0)}_{\perp,n_{\rho},m < 0})(E^{(\pm,Z >
      0)}_{\perp,n_{\rho},m < 0} \pm 2\lambda B) (E^{(\pm,Z >
      0)}_{\perp,n_{\rho},m < 0} \pm 2\lambda B \mp M)}}\nonumber\\
 \hspace{-0.3in}&& \times\left(\begin{array}{c} i ZeB\,(E^{(\pm,Z >
 0)}_{n_{\rho},m < 0} \pm E^{(\pm,Z >
 0)}_{\perp,n_{\rho},m})\,f_{n_{\rho},m}(\vec{r}_{\perp})\\ \mp p_z (E^{(\pm,Z >
 0)}_{\perp,n_{\rho},m < 0}\pm 2 \lambda B \mp M)\,f_{n_{\rho},m +
 1}(\vec{r}_{\perp})\\ i Z e B p_z\,f_{n_{\rho},m}(\vec{r}_{\perp}) \\ \pm
 (E^{(\pm,Z > 0)}_{\perp,n_{\rho},m < 0} \pm 2 \lambda B \mp M)\,(E^{(\pm,Z
 > 0)}_{n_{\rho},m < 0} \pm E^{(\pm,Z > 0)}_{\perp,n_{\rho},m < 0})\,f_{n_{\rho},m +
 1}(\vec{r}_{\perp})
  \end{array}\right)\, 
e^{-iE^{(\pm, Z > 0)}_{n_{\rho},m < 0} t + ip_z z},
\end{eqnarray} 
and negatively charged ($Z < 0$) fermions
\begin{eqnarray}\label{label30}
 \hspace{-0.3in}&& \psi^{(\pm, Z < 0)}_{n_{\rho},m \ge 0, p_z}(x) =
 \sqrt{\frac{\pi (n_{\rho} - 1)!(|Z|eB)^{m+2}}{2^{m+1}(n_{\rho} +
     m)!}}\frac{1}{\sqrt{(E^{(\pm,Z < 0)}_{n_{\rho},m \ge 0} \pm
     E^{(\pm,Z < 0)}_{\perp,n_{\rho},m \ge 0})(E^{(\pm,Z <
       0)}_{\perp,n_{\rho},m \ge 0}\pm 2 \lambda B) (E^{(\pm,Z <
       0)}_{\perp,n_{\rho},m \ge 0} \pm 2 \lambda B \mp
     M)}}\nonumber\\
 \hspace{-0.3in}&& \times\left(\begin{array}{c} 2 i (n_{\rho} + m)\,(E^{(\pm,Z < 0)}_{n_{\rho},m \ge 0} \pm E^{(\pm,Z < 0)}_{\perp,n_{\rho},m \ge
 0})\, f_{n_{\rho}-1,m}(\vec{r}_{\perp})\\ 
 \pm p_z (E^{(\pm,Z < 0)}_{\perp,n_{\rho},m \ge 0}\pm 2 \lambda B \mp M)\,f_{n_{\rho}-1,m + 1}(\vec{r}_{\perp})\\ 
 2 i (n_{\rho} + m)\, p_z\,f_{n_{\rho}-1,m}(\vec{r}_{\perp}) \\ 
 \mp (E^{(\pm,Z < 0)}_{\perp,n_{\rho},m \ge 0}\pm 2\lambda B \mp M)\,(E^{(\pm,Z < 0)}_{n_{\rho},m
 \ge 0} \pm E^{(\pm,Z < 0)}_{\perp,n_{\rho},m \ge 0})\,f_{n_{\rho}-1,m +
 1}(\vec{r}_{\perp})
  \end{array}\right)\, 
e^{-iE^{(\pm, Z < 0)}_{n_{\rho},m \ge 0} t + ip_z z},
\end{eqnarray}
and 
\begin{eqnarray}\label{label31}
\hspace{-0.3in}&& \psi^{(\pm, Z < 0)}_{n_{\rho},m < 0, p_z}(x) =
  \sqrt{\frac{\pi n_{\rho}!(|Z|eB)^{|m|}}{2^{|m|-1}(n_{\rho}+|m|-1)!}}\frac{1}{\sqrt{(E^{(\pm,Z <
  0)}_{n_{\rho},m < 0} \pm E^{(\pm,Z < 0)}_{\perp,n_{\rho},m < 0})(E^{(\pm,Z <
  0)}_{\perp,n_{\rho},m < 0}\pm 2 \lambda B) (E^{(\pm,Z < 0)}_{\perp,n_{\rho},m <
  0} \pm 2 \lambda B \mp M)}}\nonumber\\
 \hspace{-0.3in}&& \times\left(\begin{array}{c} i |Z|eB\,(E^{(\pm,Z <
     0)}_{n_{\rho},m < 0} \pm E^{(\pm,Z < 0)}_{\perp,n_{\rho},m <
     0})\,f_{n_{\rho}-1,m}(\vec{r}_{\perp})\\ 
     \mp p_z (E^{(\pm,Z <
     0)}_{\perp,n_{\rho},m < 0} \pm 2 \lambda B \mp M)\,f_{n_{\rho},m
     + 1}(\vec{r}_{\perp})\\ 
     i |Z| e B p_z\,f_{n_{\rho}-1,m}(\vec{r}_{\perp}) \\ 
     \pm (E^{(\pm,Z < 0)}_{\perp,n_{\rho},m < 0} \pm 2 \lambda B \mp M)\,(E^{(\pm,Z <
     0)}_{n_{\rho},m < 0} \pm E^{(\pm,Z < 0)}_{\perp,n_{\rho},m <
     0})\,f_{n_{\rho},m + 1}(\vec{r}_{\perp})
  \end{array}\right)\, 
e^{-iE^{(\pm, Z < 0)}_{n_{\rho},m < 0} t + ip_z z},
\end{eqnarray} 
where $n_{\rho} = 0,1,2,\ldots$ and $m = 0,\pm 1, \pm 2, \ldots$.

\subsection{3. Solution of Dirac equation for neutral  fermions with 
anomalous magnetic moments in uniform magnetic field}

For a neutral fermion with mass $M$ and an anomalous magnetic moment
$\kappa$, moving in a uniform magnetic field, the Dirac equation takes
the form
\begin{eqnarray}\label{label32}
  i\frac{\partial}{\partial
  t}\,\psi(x)=\Big(-i\vec{\alpha}\cdot\vec{\nabla} -
  2\beta\lambda\,\vec\Sigma\cdot\vec B + \beta M\Big)\psi(x),
\end{eqnarray}
where $\lambda = \kappa e/4M$. Directing a magnetic field along the
$z$--axis $\vec{B} = B\,\vec{e}_z$ and separating longitudinal and
transverse degrees of freedom relative to the magnetic field, the
Dirac equation Eq.(\ref{label32}) can be transformed into the system
of first order differential equations, which can be obtained from
Eq.(\ref{label7}) with $Z = 0$ with coupled large and small components
of the Dirac bispinor wave function, and then to the second order
differential equation with decoupled large and small components of the
Dirac wave function
\begin{eqnarray}\label{label33}
 \hspace{-0.3in}&&\Big(\frac{1}{2}\,\{\pi_+,\pi_-\} - (E_{\perp} \pm
  2\lambda B)^2 + M^2\Big)\Phi^{(\pm)}(\vec{r}_{\perp})= 0,
\end{eqnarray}
which can be obtained from Eq.(\ref{label11}) with $Z = 0$. For neutral fermions the "up" and "down" solutions coincide, i.e.
$\Phi^{(\pm)}(\vec{r}_{\perp})=\Phi^{(\pm)}_\uparrow(\vec{r}_{\perp})=\Phi^{(\pm)}_\downarrow(\vec{r}_{\perp})$.
Since for neutral fermions the canonical momentum operator is $\vec{\pi} = -
i\nabla$, in Cartesian coordinates Eq.(\ref{label32}) takes the form
\begin{eqnarray}\label{label34}
 \hspace{-0.3in}&&\Big(\bigtriangleup_{\perp} + (E_\perp \pm 2\lambda
 B)^2 - M^2\Big)\Phi^{(\pm)}(\vec{r}_{\perp}) =
 0,
\end{eqnarray}
with $\bigtriangleup_{\perp}=\partial_x^2 + \partial_y^2$. The solution of this equation can be taken in the form of a plane wave
$\Phi^{(\pm)}(\vec{r}_{\perp}) \sim
e^{\,i\vec{p}_{\perp}\cdot \vec{r}_{\perp}}$. As a result the energy
spectrum of the transverse motion of neutral fermions with anomalous
magnetic moments in a uniform magnetic field is given by
\begin{eqnarray}\label{label35}
 E^{(\pm)}_{\perp} = \sqrt{M^2 + \vec{p}^{\;2}_{\perp}} \mp 2\lambda
  B,
\end{eqnarray}
with $\vec{p}^{\;2}_{\perp}=p_x^2 + p_y^2$. The Dirac bispinor wave functions of a neutral fermion with an
anomalous magnetic moment $\kappa$, moving in a uniform magnetic
field, we search in the form
\begin{eqnarray}\label{label36}
 \hspace{-0.3in} \psi^{(\pm)}_{\vec{p}}(x) =
  \left(\begin{array}{c} F_1\\ F_2\\ F_3\\ F_4
  \end{array}\right)\, 
e^{-iE^{(\pm)}_{\vec{p}} t + i\vec{p}\cdot\vec{r}}.
\end{eqnarray}  
The wave functions Eq.(\ref{label36}) are normalised as
\begin{eqnarray}\label{label37}
\int d^3x\,\psi^{(\sigma')}_{\vec{p}'}(x)\psi^{(\sigma)}_{\vec{p}}(x)
= 2 E^{(\sigma)}_{\vec{p}}\,(2\pi)^3\,\delta^{(3)}(\vec{p}\,' -
\vec{p}\,)\,\delta_{\sigma'\sigma},
\end{eqnarray}
where $E^{(\sigma)}_{\vec{p}} = \sqrt{E^{(\sigma)2}_{\perp} + p^2_z}$
for $\sigma = \pm$. 

Substituting Eq.(\ref{label36}) into Eq.(\ref{label7}) at $Z = 0$ we
arrive at a system of algebraical equations for the normalisation
constants $F_i$ for $i = 1,2,3,4$. Solving these equations
together with the normalisation condition Eq.(\ref{label37}) we obtain
the normalised Dirac wave functions of neutral fermions
\begin{eqnarray}\label{label38}
 \hspace{-0.3in}&&\psi^{(\pm)}_{\vec{p}}(x) =
  \frac{1}{\sqrt{2(E^{(\pm)}_{\vec{p}} \pm
  E^{(\pm)}_{\perp})(E^{(\pm)}_{\perp} \pm 2\lambda B)
  (E^{(\pm)}_{\perp} \pm 2\lambda B \mp M)}} \nonumber\\
  \hspace{-0.3in}&& \times\left(\begin{array}{c} (p_x -
  ip_y)(E^{(\pm)}_{\vec{p}} \pm E^{(\pm)}_{\perp}) \\ \mp
  p_z\,(E^{(\pm)}_{\perp} \pm 2\lambda B \mp M) \\ p_z\,(p_x -
  ip_y) \\ \pm(E^{(\pm)}_{\vec{p}} \pm
  E^{(\pm)}_{\perp})(E^{(\pm)}_{\perp} \pm 2\lambda B \mp M)
  \end{array}\right)\, 
e^{-iE^{(\pm)}_{\vec{p}} t + i\vec{p}\cdot\vec{r}}.
\end{eqnarray} 
Our solutions for the relativistic wave functions of the neutral
fermions with anomalous magnetic moments, moving in a uniform magnetic
field, agree well with those, obtained in \cite{DMF2}.

\subsection{4. Non--relativistic limit of solutions of Dirac equation
 for charged and neutral fermions with anomalous magnetic moments in
uniform magnetic field}

In this section we investigate the non--relativistic limits of the
relativistic wave functions of the charged and neutral fermions and
their energy spectra.  For the positively and negatively charged
fermions in the states $(\pm ,Z > 0)$ and $(\pm ,Z < 0)$ the wave
functions take the form
\begin{eqnarray}\label{label39}
 \hspace{-0.3in}\psi^{(+, Z > 0)}_{n_{\rho},m , p_z}(x) &=& 2\pi 
 \sqrt{2M}\sqrt{\frac{ n_{\rho}!(ZeB)^{|m|+1}}{\pi(n_{\rho}+|m|)!
     2^{|m|+1}}}\left(\begin{array}{c} 1\\ 0\\ 0\\0
  \end{array}\right)\,\rho^{|m|}L^{|m|}_{n_{\rho}}
\Big(\frac{|Z|eB\rho^2}{2}\Big)\,e^{- \frac{1}{4}|Z|eB\rho^2}\,e^{i m\,\phi}\nonumber\\
\hspace{-0.3in}&&\times
e^{-i{\cal E}^{(+, Z > 0)}_{n_{\rho},m } t + ip_z z},\nonumber\\
 \hspace{-0.3in} \psi^{(-, Z > 0)}_{n_{\rho},m \ge 0, p_z}(x) &=& 2
 \pi \sqrt{2M}\sqrt{\frac{(n_{\rho} -1)!(ZeB)^{m+2}}{\pi (n_{\rho} +
     m)! 2^{m + 2}}}\left(\begin{array}{c} 0\\ 1\\ 0\\0
  \end{array}\right)\,\rho^{m+1}L^{m+1}_{n_{\rho} -1}
\Big(\frac{|Z|eB\rho^2}{2}\Big)\,e^{- \frac{1}{4}|Z|eB\rho^2}\,e^{i (m
  + 1)\,\phi}\nonumber\\
\hspace{-0.3in}&&\times e^{-i{\cal E}^{(-, Z > 0)}_{n_{\rho},m} t + ip_z
  z},\nonumber\\
\hspace{-0.3in}\psi^{(-, Z > 0)}_{n_{\rho}, m < 0, p_z}(x) &=& 2 \pi
\sqrt{2M}\sqrt{\frac{n_{\rho}!(ZeB)^{|m|}}{\pi (n_{\rho} + |m| - 1)!
    2^{|m|}}}\left(\begin{array}{c} 0\\1\\ 0\\0
  \end{array}\right)\,\rho^{|m|-1}L^{|m|-1}_{n_{\rho}}
\Big(\frac{|Z|eB\rho^2}{2}\Big)\,e^{- \frac{1}{4}|Z|eB\rho^2}\,e^{i (m + 1)\,\phi}\nonumber\\
\hspace{-0.3in}&&\times e^{-i{\cal E}^{(-, Z >
    0)}_{n_{\rho},m} t + ip_z z},
\end{eqnarray}
and 
\begin{eqnarray}\label{label40}
 \hspace{-0.3in}\psi^{(+, Z < 0)}_{n_{\rho},m , p_z}(x) &=& 2\pi
 \sqrt{2M}\sqrt{\frac{ (n_{\rho}-1)!(|Z|eB)^{|m|+1}}{\pi
     (n_{\rho}+|m|-1)!  2^{|m|+1}}}\left(\begin{array}{c} 1\\ 0\\ 0\\0
  \end{array}\right)\,\rho^{|m|}\,L^{|m|}_{n_{\rho}-1}
\Big(\frac{|Z|eB}{2}\rho^2\Big)\, e^{-\frac{1}{4}|Z|eB \rho^2}\,e^{im
  \phi} \nonumber\\
\hspace{-0.3in}&&\times e^{-i{\cal E}^{(+, Z < 0)}_{n_{\rho},m } t + ip_z
  z},\nonumber\\
\hspace{-0.3in}\psi^{(-, Z < 0)}_{n_{\rho},m \ge 0, p_z}(x) &=& 2
\pi \sqrt{2M}\sqrt{\frac{(n_{\rho}-1)!(|Z|eB)^{m + 2}}{\pi (n_{\rho}+
    m)! 2^{m + 2}}}\left(\begin{array}{c} 0\\ 1\\ 0\\0
  \end{array}\right)\,\rho^{m +1}\,L^{m + 1}_{n_{\rho}-1}\Big(\frac{|Z|eB}{2}\rho^2\Big)\,
e^{-\frac{1}{4}|Z|eB \rho^2}\,e^{i(m+1) \phi}\nonumber\\
\hspace{-0.3in}&&\times
e^{-i{\cal E}^{(+, Z < 0)}_{n_{\rho},m} t + ip_z z},\nonumber\\
\hspace{-0.3in}\psi^{(-, Z < 0)}_{n_{\rho},m < 0, p_z}(x) &=& 2 \pi
\sqrt{2M}\sqrt{\frac{ n_{\rho}!(|Z|eB)^{|m|}}{\pi (n_{\rho}+|m|-1)!
    2^{|m|}}}\left(\begin{array}{c} 0\\ 1\\ 0\\0
  \end{array}\right)\,\rho^{|m| - 1}\,L^{|m| - 1}_{n_{\rho}}\Big(\frac{|Z|eB}{2}\rho^2\Big)\, e^{-\frac{1}{4}|Z|eB \rho^2}\,e^{i(m+1)
  \phi}\nonumber\\
\hspace{-0.3in}&&\times e^{-i{\cal E}^{(+, Z < 0)}_{n_{\rho},m} t + ip_z z},
\end{eqnarray}
where $n_{\rho} = 0,1,2,\ldots$ and $m = 0, \pm 1, \pm 2, \ldots$. The
wave functions Eq.(\ref{label39}) and Eq.(\ref{label40}) are defined
up to unessential phase factors. 

In the non--relativistic approximation the energy spectra ${\cal
  E}^{(\pm, Z \gtrless 0)}_{n_{\rho},m } = E^{(\pm, Z \gtrless
  0)}_{n_{\rho},m } - M$ of charged fermions in a uniform magnetic
field can be written in the form
\begin{eqnarray}\label{label41}
 \hspace{-0.3in}&& {\cal E}^{(+, Z > 0)}_{n_{\rho},m} = \frac{Z
   eB}{M}\Big(n_{\rho} + \frac{|m| - m + 1}{2}\Big) -
 \frac{1}{2}\,\mu\,\frac{ZeB}{M} + \frac{p^2_z}{2 M},\nonumber\\
 \hspace{-0.3in}&& {\cal E}^{(-, Z > 0)}_{n_{\rho},m \ge 0} = \frac{Z
   eB}{M}\Big(n_{\rho} - 1 + \frac{(|m| + 1) - (m + 1) + 1}{2}\Big) +
 \frac{1}{2}\,\mu\,\frac{ZeB}{M} + \frac{p^2_z}{2 M},\nonumber\\
\hspace{-0.3in}&& {\cal E}^{(-, Z > 0)}_{n_{\rho},m < 0} = \frac{Z
  eB}{M}\Big(n_{\rho} + \frac{(|m| - 1) - (m + 1) + 1}{2}\Big) +
\frac{1}{2}\,\mu\,\frac{ZeB}{M} + \frac{p^2_z}{2 M},\nonumber\\
 \hspace{-0.3in}&& {\cal E}^{(+, Z < 0)}_{n_{\rho},m} = \frac{|Z|
   eB}{M}\Big(n_{\rho} - 1 + \frac{|m| + m + 1}{2}\Big) +
 \frac{1}{2}\,\mu\,\frac{|Z|eB}{M}+ \frac{p^2_z}{2 M},\nonumber\\
 \hspace{-0.3in}&& {\cal E}^{(-, Z < 0)}_{n_{\rho},m \ge 0} = \frac{|Z|
   eB}{M}\Big(n_{\rho} - 1 + \frac{(|m| + 1) + (m + 1) + 1}{2}\Big) -
 \frac{1}{2}\,\mu\,\frac{|Z|eB}{M}+ \frac{p^2_z}{2 M},\nonumber\\
 \hspace{-0.3in}&& {\cal E}^{(-, Z < 0)}_{n_{\rho},m < 0} = \frac{|Z|
   eB}{M}\Big(n_{\rho} + \frac{(|m| - 1) + (m + 1) + 1}{2}\Big) -
 \frac{1}{2}\,\mu\,\frac{|Z|eB}{M}+ \frac{p^2_z}{2 M}.
\end{eqnarray}
The terms, proportional to a total magnetic moment $\mu = 1 + \kappa$,
correspond to the Pauli energy splitting \cite{Landau3}. The
non--relativistic energy spectra Eq.(\ref{label41}) correspond to the
quantum states, described by the non--relativistic wave functions
Eq.(\ref{label39}) and Eq.(\ref{label40}) of positively and negatively
charged fermions.

Using the properties of the Laguerre polynomials \cite{Lebedew} we may
make a change of the quantum numbers $n_{\rho} - 1 \to n_{\rho}$ for the second 
wave function of Eq.(\ref{label39}) as well as the first and second one of Eq.(\ref{label40}).
Additionally, all wave functions with polarisation {\it down} are shifted $m + 1 \to m$. This brings the
non--relativistic wave functions of positively and negatively charged
fermions in Eq.(\ref{label39}) and Eq.(\ref{label40}) to a unified form
\begin{eqnarray}\label{label42}
 \hspace{-0.3in}\psi^{(\pm, Z \gtrless 0)}_{n_{\rho},m , p_z}(x) =
  2\pi \sqrt{2M}\sqrt{\frac{
     n_{\rho}!(|Z|eB)^{|m|+1}}{\pi(n_{\rho}+|m|)!
     2^{|m|+1}}}\,\varphi_{\pm}\,\rho^{|m|}L^{|m|}_{n_{\rho}}
 \Big(\frac{|Z|eB\rho^2}{2}\Big)\,e^{- \frac{1}{4}|Z|eB\rho^2}\,e^{i
   m\,\phi} e^{-i{\cal E}^{(+, Z \gtrless 0)}_{n_{\rho},m } t + ip_z
   z},
 \hspace{-0.3in} 
\end{eqnarray}
agreeing fully with the non--relativistic wave functions obtained in
\cite{Landau3}, $\varphi_{\pm}$ are the wave functions with elements 
$\varphi_+ = (1,0,0,0)$ and $\varphi_- = (0,1,0,0)$, respectively.  These wave functions describe the
non--relativistic quantum states with the energy spectra
\begin{eqnarray}\label{label43}
 \hspace{-0.3in}&& {\cal E}^{(\pm, Z > 0)}_{n_{\rho},m} = \frac{Z
   eB}{M}\Big(n_{\rho} + \frac{|m| - m + 1}{2}\Big) \mp
 \frac{1}{2}\,\mu\,\frac{ZeB}{M} + \frac{p^2_z}{2 M},\nonumber\\
 \hspace{-0.3in}&& {\cal E}^{(\pm, Z < 0)}_{n_{\rho},m} = \frac{|Z|
   eB}{M}\Big(n_{\rho} + \frac{(|m| + m + 1}{2}\Big) \pm 
 \frac{1}{2}\,\mu\,\frac{|Z|eB}{M}+ \frac{p^2_z}{2 M},
\end{eqnarray}
for positively and negatively charged fermions, respectively, with
quantum numbers $n_{\rho} = 0,1, \ldots$ and $m = 0,\pm 1,
\ldots$. 

In the non--relativistic limit the wave functions of a neutral fermion
with an anomalous magnetic moment $\kappa$ are equal to
\begin{eqnarray}\label{label44}
 \hspace{-0.3in}\psi^{(+)}_{\vec{p}}(x) =
 \sqrt{2M}\left(\begin{array}{c} 1\\0\\0\\0 
  \end{array}\right)\, 
 e^{-i{\cal E}^{(+)}_{\vec{p}} t + i\vec{p}\cdot \vec{r}}\quad,\quad
 \psi^{(-)}_{\vec{p}}(x) = \sqrt{2 M}\left(\begin{array}{c} 0\\ 1 \\
 0\\ 0
  \end{array}\right)\, 
e^{-i{\cal E}^{(-)}_{\vec{p}} t + i\vec{p}\cdot \vec{r}}.
\end{eqnarray} 
The energy spectra ${\cal E}^{(\pm )}_{\vec{p}}$ are given by
\begin{eqnarray}\label{label45}
 {\cal E}^{(\pm)}_{\vec{p}} = \frac{\vec{p}^{\;2}}{2 M}\mp
 \frac{1}{2}\kappa \frac {eB}{M}.
\end{eqnarray} 
The Pauli energy splitting is defined by the last term in
Eq.(\ref{label44}) \cite{Landau3}. 

\subsection{5. Quantum states of charged fermions with anomalous 
magnetic moments in uniform magnetic field with radial quantum number
$n_{\rho} = 0$}

The analysis of quantum states with zero quantum numbers is one of the
most important mathematical problems of Quantum mechanics \cite{Landau3}
as a part of Mathematical physics \cite{Lebedew}. In section 2 we have
found the relativistic wave functions and energy spectra of charged
fermions with anomalous magnetic moments coupled to a uniform magnetic
field. These wave functions have a rather complicated dependence on
the radial quantum number $n_{\rho}$, which does not make obvious the
existence of the correct quantum states for $n_{\rho} = 0$. In this
section, skipping rather tedious and cumbersome intermediate
calculation we adduce the wave functions of quantum states of charged
fermions with anomalous magnetic moments, moving in a uniform magnetic
field with radial quantum number $n_{\rho} = 0$. 

First, we consider the quantum states described by the wave functions
Eq.(\ref{label28}) and Eq.(\ref{label31}), the energy spectra of which
do not depend on the magnetic quantum number. Taking the wave
functions Eq.(\ref{label28}) and Eq.(\ref{label31}) in the limit
$n_{\rho} \to 0$ with arbitrary magnetic quantum numbers $m$ and
skipping intermediate calculations we obtain the following expressions
\begin{eqnarray}\label{label46}
\hspace{-0.3in}&&\psi^{(+, Z > 0)}_{0,m \ge 0, p_z}(x) =
i\sqrt{\frac{\pi (ZeB)^{m+1}}{2^{m-1} m!}}\sqrt{E^{(+,Z > 0)}_{0,m \ge
    0} + E^{(+,Z > 0)}_{\perp,0,m \ge 0}}\left(\begin{array}{c}
  f_{0,m}(\vec{r}_{\perp})\\ 0\\ {\displaystyle \frac{p_z}{E^{(+,Z >
        0)}_{0,m \ge 0} + E^{(+,Z > 0)}_{\perp,0,m \ge
        0}}}\,f_{0,m}(\vec{r}_{\perp}) \\ 0
  \end{array}\right)\, 
e^{-iE^{(+, Z > 0)}_{0,m \ge 0} t + ip_z z},\nonumber\\ 
\hspace{-0.3in}&&\psi^{(-, Z >
  0)}_{0,m \ge 0, p_z}(x) =  0, \nonumber\\ 
\hspace{-0.3in}&&\psi^{(+, Z < 0)}_{0,m <
  0, p_z}(x) =  0,\nonumber\\ 
\hspace{-0.3in}&&\psi^{(-, Z < 0)}_{0,m < 0, p_z}(x) = \sqrt{\frac{\pi
    (|Z|eB)^{|m|}}{2^{|m|-2}(|m|-1)!}}\sqrt{E^{(-,Z < 0)}_{0,m < 0} +
  E^{(-,Z < 0)}_{\perp,0,m < 0}} \left(\begin{array}{c}0\\ f_{0,m +
    1}(\vec{r}_{\perp})\\0 \\ - {\displaystyle \frac{p_z}{E^{(-,Z <
        0)}_{0,m < 0} + E^{(-,Z < 0)}_{\perp,0,m < 0}}}\,f_{0,m +
    1}(\vec{r}_{\perp})
  \end{array}\right)\, 
e^{-iE^{(-m, Z < 0)}_{0,m < 0} t + ip_z z}.\nonumber\\
\hspace{-0.3in}&&
\end{eqnarray}
The quantum states of positively charged fermions, described by the
wave functions Eq.(\ref{label29}), depend on the radial quantum number
$n_{\rho}$ and the magnetic quantum number $m < 0$, i.e. $m = -1,-2,
\ldots$. Setting $n_{\rho} = 0$ we obtain
\begin{eqnarray}\label{label47}
\hspace{-0.3in}&& \psi^{(\pm, Z > 0)}_{0,m < 0, p_z}(x) =
\sqrt{\frac{\pi (ZeB)^{|m|}}{2^{|m| -
      1}(|m|-1)!}}\frac{1}{\sqrt{(E^{(\pm,Z > 0)}_{0,m
      < 0} \pm E^{(\pm,Z > 0)}_{\perp,0,m < 0})(E^{(\pm,Z >
      0)}_{\perp,0,m < 0} \pm 2\lambda B) (E^{(\pm,Z >
      0)}_{\perp,0,m < 0} \pm 2\lambda B \mp M)}}\nonumber\\
 \hspace{-0.3in}&& \times\left(\begin{array}{c} i ZeB\,(E^{(\pm,Z >
     0)}_{0,m < 0} \pm E^{(\pm,Z > 0)}_{\perp,0,m <
     0})\,f_{0,m}(\vec{r}_{\perp})\\ \mp p_z (E^{(\pm,Z >
     0)}_{\perp,0,m < 0}\pm 2 \lambda B \mp M)\,f_{0,m +
     1}(\vec{r}_{\perp})\\ i Z e B p_z\,f_{0,m}(\vec{r}_{\perp})
   \\ \pm (E^{(\pm,Z > 0)}_{\perp,0,m < 0} \pm 2 \lambda B \mp
   M)\,(E^{(\pm,Z > 0)}_{0,m < 0} \pm E^{(\pm,Z > 0)}_{\perp,0,m <
     0})\,f_{0,m + 1}(\vec{r}_{\perp})
  \end{array}\right)\, 
e^{-iE^{(\pm, Z > 0)}_{0,m < 0} t + ip_z z}.
\end{eqnarray}
The wave functions Eq.(\ref{label30}) describe the quantum states of
negatively charged fermions with anomalous magnetic moments,
characterised by the radial quantum number $n_{\rho}$ and the
magnetic quantum number $m \ge 0$, i.e. $m = 0, +1, +2, \ldots$. The
quantum states, characterised by the quantum numbers $n_{\rho} = 0$, $m \geq 0$
are defined by the wave functions
\begin{eqnarray}\label{label48}
 \psi^{(\pm, Z < 0)}_{0,m \geq 0, p_z}(x) = 0.
\end{eqnarray}
The properties of the relativistic wave functions obtained at
$n_{\rho} = 0$ q, taken in the non--relativistic limit, agree well
with the properties of the non--relativistic wave functions, given in
Eq.(\ref{label39}) and Eq.(\ref{label40}).

\subsection{6. Conclusive discussion}

We have proposed a solution of the Dirac equation for charged and
neutral fermions with spin $\frac{1}{2}$ and anomalous magnetic
moments, moving in a uniform magnetic field. It is well--known that
the Dirac equation, describing the motion of a relativistic fermion with
spin $\frac{1}{2}$ in any external field, can be transformed into a
system of first order differential equations for the wave functions of
the coupled {\it up} and {\it down} spin states of the large and small
components of the Dirac wave function \cite{IZ80}. The procedure of a
disentanglement of the {\it up} and {\it down} spin states of the
large and small components of the Dirac wave function depends on the
structure of the external field and, of course, the properties of the
fermions. For charged fermions without anomalous magnetic moment the
differential equations, describing the wave functions of the
disentangled {\it up} and {\it down} spin states, are of second order
\cite{DMF1,DMF3}--\cite{DMF6}. As we have shown above, a non-vanishing
anomalous magnetic moment leads to fourth order differential equations
for the disentangled {\it up} and {\it down} spin states. This agrees
well with the analysis of the energy spectra of charged fermions with
spin $\frac{1}{2}$ and anomalous magnetic moment carried out in
\cite{DMF6}. These fourth order differential equations can be reduced
to second order ones with eigenvalues yielding the Pauli energy
splitting. The wave functions of the disentangled {\it up} and {\it
  down} spin states are calculated in cylindrical coordinates in terms
of the generalised Laguerre polynomials $L_{n_{\rho}}^{|m|}$ and
circular functions $e^{im\phi}$, depending on the radial quantum
number $n_{\rho} = 0,1,2,\ldots$ and the magnetic quantum number $m =
0,\pm 1,\pm 2, \ldots$. According to \cite{DMF9}, only the solutions
of the Dirac equation in cylindrical coordinates for charged fermions,
moving in a uniform magnetic field, can adequately describe a
two--dimensional motion of fermions in the plane orthogonal to a
uniform magnetic field.

For the calculation of the normalised Dirac wave functions we use the
system of first order differential equations for the wave functions
Eq.(\ref{label7}) of the coupled {\it up} and {\it down} spin states
and the normalisation conditions Eq.(\ref{label26})
and Eq.(\ref{label37}) for charged and neutral fermions,
respectively. In the limit of vanishing anomalous magnetic moment,
$\kappa = 0$, the obtained Dirac wave functions coincide with those,
given in \cite{DMF1,DMF2}--\cite{DMF6}. In comparison with the wave
functions, obtained in \cite{DMF2}, the solutions, proposed above, are
more detailed and convenient for applications.

In the non--relativistic limit the wave functions of the charged
fermions coincide with the well--known solutions of the Schr\"odinger
equation \cite{Landau3}. The relativistic wave functions of a neutral
fermion with an anomalous magnetic moment, moving in a uniform
magnetic field, have a shape of plane waves with an energy spectrum,
splitted by the interaction of its anomalous magnetic moment with a
uniform magnetic field.

The relativistic energy spectra of charged fermions in a uniform
magnetic field are calculated in dependence of the radial $n_{\rho} =
0,1,2,\ldots$ and magnetic $m = 0,\pm 1, \pm 2,\ldots$ quantum
numbers. All of these energy levels have an additional splitting,
which is caused by the interaction of the anomalous magnetic moments of
fermions with a uniform magnetic field. The energy spectra of
positively charged fermions with positive magnetic quantum number and
negatively charged fermions with negative magnetic quantum numbers are
infinitely degenerated \cite{DMF9}.  In the non--relativistic limit
the energy spectra reproduce the well--known Landau energy spectra
with the Pauli energy splitting \cite{Landau3}.

Of course, the wave functions and the energy spectra of charged
fermions, given in terms of the radial $n_{\rho}$ and magnetic $m$
quantum numbers, can be presented in terms of the principal quantum
number $n$, expressed in terms of the radial and magnetic quantum
numbers, and the magnetic quantum number $m$. However, such a
definition of the fermion states in a uniform magnetic field is not
convenient for practical calculations.

Comparing our results with those, obtained in
\cite{DMF6}--\cite{DMF8}, we would like to accentuate that our
technique of the solution of the Dirac equation is similar to that
used in \cite{DMF6} but simpler. In addition to the energy spectra,
which were obtained in \cite{DMF6}, we have calculated the relativistic
wave functions. As regards the results, obtained in \cite{DMF7,DMF8},
where the solutions of the Dirac equation were found
in the ``number of states'' representation, we have calculated not
only the wave functions and energy spectra, but also investigated a
detailed dependence of the relativistic wave functions on the magnetic
quantum number and analysed their non--relativistic limits. A detailed
behaviour of the relativistic wave functions and their exact
dependence on the quantum numbers play an important role for
applications of the obtained results, for example, to the neutron
$\beta^-$--decay.

This work was supported by the Austrian ``Fonds zur F\"orderung der
Wissenschaftlichen Forschung'' (FWF) under contract I689-N16 and
contract I862-N20 AXION and in part by the U.S. Department of Energy
contract no. DE-FG02-08ER41531, no. DE-AC02-06CH11357 and by the
Wisconsin Alumni Research Foundation.


\begin{thebibliography}{99}
\bibitem{DMF1}
M. H. Johnson, B. A. Lippmann,
Phys. Rev. {\bf 76}, 828 (1949), Phys. Rev. {\bf 77}, 702 (1950).
\bibitem{IZ80}
C. Itzykson and J.--B. Zuber,
in {\it Quantum Field Theory}, McGraw--Hill Inc., 1980.
\bibitem{DMF2}
J. J. Matese, R. F. O'Connell,
Phys. Rev. {\bf 180}, 1289 (1969).
\bibitem{DMF3}
L. Fassio--Canuto,
Phys. Rev. {\bf 187}, 2141 (1969).
\bibitem{DMF4}
V. L. Kauts, A. M. Savochkin, and A. I. Studenikin,
Phys. Atom. Nucl. {\bf 69}, 1453 (2006).
\bibitem{DMF5}
K. A. Kouzakov, A. I. Studenikin,
Phys. Rev. C {\bf 72}, 015502 (2005).
\bibitem{DMF6}
W.-Y. Tsai and A. Yildiz,
Phys. Rev. D {\bf 4}, 3643 (1971).
\bibitem{DMF7}
M. Seetharaman, J. Prabhakaran, and P. M. Mathews,
Phys. Rev. D {\bf 12}, 458 (1975).
\bibitem{DMF8}
R. Chand, G. Szamosi,
Nuovo Cimento Letters, {\bf 22}, 660 (1978).
\bibitem{DMF9}
P. M. Mathews,
Phys. Rev. D {\bf 9}, 365 (1974).
\bibitem{Landau3}
L. D. Landau and E. M. Lifschitz, {\it Quantenmechanik},
Verlag Harri Deutsch, (2007).
\bibitem{Lebedew} 
N. N. Lebedew, in {\it Spezielle Fuktionen und 
ihre Anwendungen}, Wissenschaftsverlag, Mannheim, (1973).
\end{thebibliography}
\end{document}